\documentstyle[prl,aps,epsf]{revtex}
\twocolumn
\tighten

\begin{document}
\draft
\twocolumn[
\hsize\textwidth\columnwidth\hsize\csname@twocolumnfalse\endcsname
\title{Correlation between magnetic and transport properties of phase separated
La$_{0.5}$Ca$_{0.5}$MnO$_{3}$}
\author{J. Sacanell$^{a}$, P. Levy$^{a}$, L. Ghivelder$^{b}$,
G. Polla$^{a}$, F. Parisi$^{a}$}
\address{
(a) Departamento de F\'{\i}sica, Comisi\'{o}n Nacional de Energ\'{\i}a At\'{o}mica,
Avda. Gral Paz 1499 (1650) San Mart\'{\i}n, Buenos Aires, Argentina  \\
(b) Instituto de F\'{\i}sica, Universidade Federal do Rio de
Janeiro, C.P. 68528, Rio de Janeiro, RJ 21945-970, Brazil
}

\maketitle

\begin{abstract}
The effect of low magnetic fields on the magnetic and electrical transport properties of
polycrystalline samples of the phase separated compound La$_{0.5}$Ca$_{0.5}$MnO$_{3}$ is studied.
The results are interpreted in the framework of the field induced ferromagnetic fraction
enlargement mechanism. A fraction expansion coefficient $\alpha_{f}$, which relates the
ferromagnetic fraction $f$ with the applied field $H$, was obtained. A phenomenological model to
understand the enlargement mechanism is worked out.\\
\end{abstract}
]
\narrowtext

\section{Introduction}

The relation between the phase separation (PS) phenomena (the coexistence of two or more phases)
occurring in some manganites (manganese oxides) and the colossal magnetoresistance ($CMR$) effect
is being the focus of extensive research since the work of Uehara et al. \cite{uehara} showing
the crucial interplay between both phenomena.\\
Recently, it has been shown that  the $CMR$ effect observed in phase separated systems can be
accounted for by a single mechanism called "field induced ferromagnetic fraction enlargement"
(FIFFE) \cite{parisi}. The main hypothesis lying in it is that by applying a low magnetic field
in a field-cooled experiment it is possible to affect the relative fraction of the coexisting
phases, giving rise to qualitative and quantitative changes in the physical properties of the
material. This mechanism has been also considered as the source of the $CMR$ effect in a recent
theoretical work \cite{mayr}.\\
The aim of this paper is to propose a quantitative model for the FIFFE mechanism. To do this we
have performed  magnetization and transport measurements under low magnetic fields (between 0 and
1 Tesla) on the prototypical PS compound La$_{0.5}$Ca$_{0.5}$MnO$_{3}$.\\
La$_{0.5}$Ca$_{0.5}$MnO$_{3}$, paramagnetic at room temperature, shows a mainly ferromagnetic
(FM) state close below $T_{C}($225 K and a charge ordered antiferromagnetic (COAF) state for
$T<T_{co}\approx$150 K on cooling. However, at low temperature, a fraction of the FM phase
remains trapped in the COAF host, giving rise to magnetic and transport responses characteristic
of the PS state. The amount of the low temperature FM phase ($f$) and its spatial distribution
were shown to be highly influenced by the sample preparation procedures, being the ceramic grain
size one of the ways of controlling them. \cite{levy}  This fact gives us the unique possibility
to study the magnetic field effects on samples with different zero field FM fractions ($f(0)$),
keeping almost unchanged characteristics parameters as $T_{C}$ and $T_{co}$.\\

\section{Experimental}

Polycrystalline samples of La$_{0.5}$Ca$_{0.5}$MnO$_{3}$ of several grain sizes were used, their
preparation procedure and structural characterization are described elsewhere \cite{levy}.
Transport measurements were performed with the standard four probe method. Magnetization
measurements were performed in a commercial magnetometer (Quantum Design PPMS).\\

\section{Results}

In Fig. 1a we show the resistivities of the samples under study. Samples I and II have
$f(0)\approx$ 31 \% and $f(0)\approx$ 11\% (obtained through magnetization measurements) at low
temperature. They display metallic like behavior below 80 K indicating the presence of
percolative paths of the FM phase. Sample III ($f(0)\approx$ 9 \%) has an insulating behavior and
no signal of percolation was found on cooling down to 40 K. Previous experiments indicate that
$f(0)$ for this sample is close below the percolation limit. Magnetoresistance ($MR$) curves,
obtained upon cooling these samples in the presence of a field $H=$0.6 T, are displayed in Figs
1b to 1d. The overall $MR$ values increase from one sample to other as the low temperature FM
content decrease. Samples I and II have $f(0) > f_{c}$ (percolation limit) and so, there's no
qualitative difference between them but only a quantitative one. Their different low $T$ $MR$
values can only be accounted for by the FIFFE mechanism. On the other hand, sample III has $f(0)$
close below$f_{c}$ and its colossal MR value at low $T$ can be explained by the increment of $f$,
which induces a change of regime from non percolative to percolative.\\

\begin{center}
\begin{figure}[tbp]
\hspace{-2cm}
\epsfxsize=5.50cm
\centerline{\epsffile{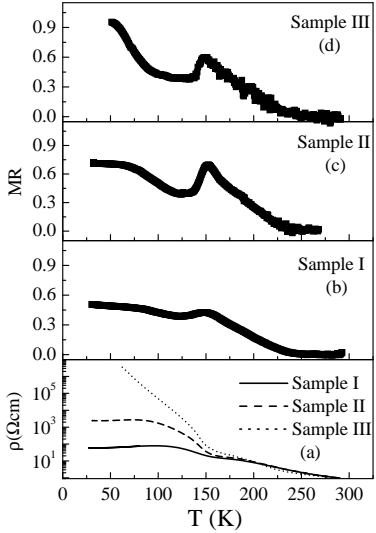}}
\caption{ Resistivity (a) and Magnetoresistance (b to d) of samples I, II and III, on cooling
under $H$=0.6 Tesla.
\label{Fig1}
}
\end{figure}
\end{center}

In order to give insight into the field dependence of $f$ we have performed magnetization and
resistivity measurements as a function of the cooling field $H_{c}$ on the above described
samples. To obtain this information from magnetization measurements we have used the following
procedure: the sample was firstly cooled in presence of $H_{c}$ (with $H_{c}$ between 0 and 1 T)
down to 30 K. At this temperature a measuring field $H_{m}=$1 T was applied for the acquirement
of the magnetization value $M(H_{c}, H_{m})$. Following the fact that this low measuring field
does not affect the relative fractions of the coexisting phases when applied at $T << T_{co}$, we
can obtain the $H_{c}$ dependence $f$ at T=30 K as $f(H_{c})=M (H_{c}, H_{m})/M_{hom}(H_{m})$,
where $M_{hom}(H_{m})($3.5 (B/Mn is the magnetization of an homogeneous FM sample at $H_{m}=$1
T.\\
The results for samples I and III are sketched in Fig. 2a. We can see that for $H$ greater than
0.2 Tesla $f(H)$ follows a linear dependence $f(H)=f(0)+ \alpha_f H$, where $\alpha_f$, named the
fraction expansion coefficient, is the fundamental parameter relating $f$ with $H$.\\

\begin{center}
\begin{figure}[tbp]
\hspace{-2cm}
\epsfxsize=5.50cm
\centerline{\epsffile{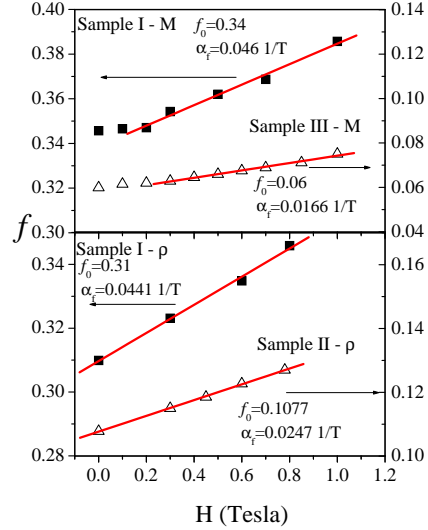}}
\caption{Magnetic field dependence of the FM fraction obtained through: (a) magnetization and (b)
resistivity experiments. The data points were adjusted with $f_{0} + \alpha_f H$.
\label{Fig2}
}
\end{figure}
\end{center}

An indirect determination of $f(H)$ can be obtained from resistivity measurements using a semi
phenomenological model for transport through a binary mixture known as General Effective Medium
theory or GEM \cite{mclach}, which provides a relation to obtain the samples' resistivity
$\rho_{e}$ as a function of $\rho_{FM}$ and $\rho_{co}$, the resistivities of the constitutive
phases:
\begin{equation}
f\frac
{(\rho_{e}^{1/t}-\rho_{FM}^{1/t})}{(\rho_{e}^{1/t}-A_c\rho_{FM}^{1/t})}+
(1-f)\frac{(\rho_{e}^{1/t}-\rho_{co}^{1/t})}{(\rho_{e}^{1/t}-A_c\rho_{co}^{1/t})
}
=
0
\end{equation}
where $A_c=1/f_c-1$ depends on the critical percolation fraction $f_{c}$ and $t\approx$2 depends
on the geometry of the regions. \\
The resistivity of the three samples was measured  while cooling them under several magnetic
fields (0, 0.3, 0.45, 0.6 and 0.75 Tesla). The FM fraction $f(H)$ was calculated at 30 K using
the GEM expression. The calculus was done taking into account the variation of $\rho_{FM}$ given
by the application of a magnetic field, due to the "normal" {\it MR} effect on homogeneous
samples. We have assumed that $\rho_{co}$ does not change with the application of a low $H$. In
Fig. 2b we show the so obtained data for samples I and II, reflecting the same linear dependence
between $f$ and $H$ that was observed through magnetization measurements.\\

\section{Discussion}

The analysis of the fitting parameters obtained by both magnetization and resistivity methods,
shows an interplay between $f(0)$ and $\alpha_f$. In Fig. 3 we sketch the dependence of ($log
\alpha_f$) with ($log f(0)$) obtained from experimental curves like those of Fig. 2. On one hand,
it is worth noting that both determination methods give similar values of $f(0)$ and ($\alpha_f$)
for all samples. This fact gives some confidence on the phenomenological model used to describe
the resistivity which, as was stated above, is an indirect model-dependent method. On the other
hand, the three independent points in Fig. 3 seem to indicate that there is a linear relation
between ($log \alpha_f$) and ($log f(0)$), then a power law $\alpha_f\approx f(0)^{z}$
($z\approx$ 0.53) is found. Under this hypothesis a geometrical view for the FIFFE mechanism can
be developed. Supposing that $r$ is a typical spatial parameter to describe the FM content of a
sample ($f\approx r^{n}$ with  $n=$3 if the FM regions are described as spheres, and $n=$2 if
they are rather filaments) and that the application of a field $H$ yields the increase of $r$ in
$\Delta r$ (i.e., $\Delta f\approx r^{n-1}$) the relation $\alpha_f\approx f^{z}$ with $z=1-1/n$
is obtained. The experimental result ($z=$0.53) is then consistent with the filamentary view
which, in turn, is in close agreement with the fact that the FM phase is distributed along
percolative or nearly percolative paths in the studied samples.\\

\begin{center}
\begin{figure}[tbp]
\hspace{-2cm}
\epsfxsize=5.50cm
\centerline{\epsffile{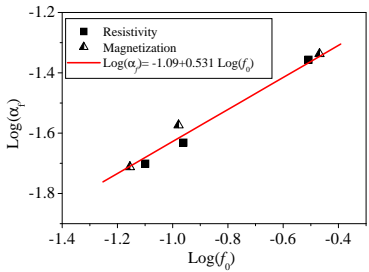}}
\caption{FM fraction expansion coefficient vs. $f_{0}$ for all experiments.
\label{Fig3}
}
\end{figure}
\end{center}

\section{Concluding Remarks}

In conclusion, the study of $f$ vs. $H$  gives us a direct image of the FIFFE  mechanism, which
seems to be the responsible for low field {\it CMR} in PS systems \cite{parisi}. The results
presented here show, for low $H$, a linear dependence between $f$ and the applied field.
Particularly, the slope of the $f$ vs. $H$ curves, $\alpha_f$, arises as a very important
parameter to describe the interplay between magnetic and transport properties in PS systems. We
also worked out a simple geometrical model to understand the dependence of $\alpha_f$with the
initial FM fraction $f(0)$, which is sample dependent. In this model the FIFFE effect  is mainly
determined by the growth of the FM phase against the COAF one through the interface between them.
Further experimental and theoretical  work is needed to elucidate this point.\\

\end{document}